\documentclass[traditabstract]{aa}
\usepackage[varg]{txfonts}
\usepackage{graphicx}
\usepackage{journals}
\usepackage{natbib}
\bibpunct{(}{)}{;}{a}{}{,}

\def\fracd#1#2{{\displaystyle\frac{#1}{#2}}}

\begin{document}
\title{A new approach to the assessment of stochastic errors of radio source position catalogues}
\author{Zinovy Malkin$^{1,2}$}
\institute{$^1$Pulkovo Observatory, St.~Petersburg 196140, Russia\\
  $^2$St. Petersburg State University, St.~Petersburg 198504, Russia}
\date{Received / Accepted}
\titlerunning{Stochastic errors assessment of source position catalogues}
\authorrunning{Z. Malkin}
\abstract{
Assessing the external stochastic errors of radio source position catalogues derived from VLBI observations is important for tasks such
as estimating the quality of the catalogues and their weighting during combination.
One of the widely used methods to estimate these errors is the three-cornered-hat technique, which can be extended to the N-cornered-hat technique.
A critical point of this method is how to properly account for the correlations between the compared catalogues.
We present a new approach to solving this problem that is suitable for simultaneous investigations of several catalogues.
To compute the correlation between two catalogues $A$ and $B$, the differences between these catalogues and a third arbitrary catalogue $C$
are computed.
Then the correlation between these differences is considered as an estimate of the correlation between catalogues $A$ and $B$.
The average value of these estimates over all catalogues $C$ is taken as a final estimate of the target correlation.
In this way, an exhaustive search of all possible combinations allows one to compute the paired correlations between all catalogues.
As an additional refinement of the method, we introduce the concept of weighted correlation coefficient.
This technique was applied to nine recently published radio source position catalogues.
We found large systematic differences between catalogues, that significantly impact determination of their stochastic errors.
Finally, we estimated the stochastic errors of the nine catalogues.
}
\maketitle


\section{Introduction}

Very long baseline interferometry (VLBI) is currently the primary technique for maintening International Celestial Reference Frame (ICRF, \citet{Ma2009p}).
The latter is realized as a catalogue of radio source coordinates derived from processing the VLBI observations.
Assessing the systematic and stochastic errors of radio source position catalogues (RSPCs) plays an important role in improvement of the ICRF.
The internal stochastic error of the RSPCs is determined by the source position uncertainties given in the catalogue.
The external (``absolute'') catalogue stochastic error can be assessed only from mutual comparison of several RSPCs.
The principal difficulty in determining the external errors is that we can investigate only the {\it differences} between catalogues.

\citet{Malkin2008j} proposed a method for evaluating the overall accuracy of RSPCs without separating catalogue errors into
systematic and stochastic parts based on comparison of the uncertainty of the nutation angles derived from VLBI observations using different RSPCs.
To estimate the stochastic errors only, the so-called three-cornered-hat (TCH) method can be used.
It was originally developed for investigation of the stability of frequency standards \citep{Gray1974}, and was then applied for noise analyses
of various data, in particular, astronomical and geodetic time series and RSPCs.
However, although this method is widely used, its application is not straightforward because it generally requires knowledge of the correlation
between series under investigation, which are never known a priori.

One method to overcome this difficulty was proposed by \citet{Tavella1994} and was further advanced by \citet{Torcaso1998} and \citet{Galindo2000}.
However, this method can be only partly applied to the RSPCs analysis because it requires selecting one of the compared clocks (data sets)
as a reference clock with which other clocks (data sets) are compared, which contradicts our goal of comparing all the catalogues as equipollent.
This means that the first problem is that the result will depend on the choice of the reference catalogue.
It is also important that this method is not aimed at estimating the actual correlation between data sets, but at finding the smallest correlations
that provide a positive variance solution \citep{Torcaso1998}.

Several developments in using the TCH method for RSPCs were made in the Main Astronomical Observatory (MAO) of the National Academy of Sciences,
Ukraine, which were reported in \citet{Molotaj1998} and \citet{Bolotin2010}.
To estimate the correlations between three catalogues, these authors first computed an averaged (combined) catalogue.
The differences between the input and averaged catalogues were then analyzed to derive the correlations between the three compared catalogues and
their external errors.
This method was, in particular, used to analyze of the catalogues computed in the framework of the ICRF2 project \citep{Ma2009p}.
The MAO method also has some shortcomings, partially noted by \citet{Bolotin2010}.
In particular, the results depend on the method used to compute the average catalogue and some other factors.
Moreover, this approach is suitable for comparisons of three catalogues only.

In this work, we develop a new approach to estimate the correlations between RSPCs, which is an extension of the MAO method.
It allows to simultaneously analyze an unlimited number of RSPCs, the more the better, in fact.
Another development is a new concept of weighted correlation coefficient, which is important for analysis of unevenly weighted data.
The third improvement is accounting for systematic differences between catalogues.

The outline of the paper is the following:
The basic theory of the TCH method is given in Section~\ref{sect:TCH}.
In Section~\ref{sect:application}, the proposed approach is described and applied to nine recently published RSPCs.
In particular, a new method for estimating correlation between catalogues is considered in this section.
Finally, we we estimated the stochastic errors of the nine RSPCs.
Section~\ref{sect_conclusions} provides a summary of our results.


\section{Basics of the TCH method}
\label{sect:TCH}

In its original formulation, the TCH method was applied to three series of measurements, which allows one to write the following system of three equations
for the paired differences between the series assuming they are uncorrelated:
\begin{equation}
\begin{array}{l}
\sigma^2_{12} = \sigma^2_1 + \sigma^2_2 \,, \\[1.5ex]
\sigma^2_{13} = \sigma^2_1 + \sigma^2_3 \,, \\[1.5ex]
\sigma^2_{23} = \sigma^2_2 + \sigma^2_3 \,, \\
\end{array}
\label{eq:TCH_system}
\end{equation}
with the solution
\begin{equation}
\begin{array}{l}
\sigma^2_1 =  ( \sigma^2_{12} + \sigma^2_{13} - \sigma^2_{23} ) / 2 \,, \\[2ex]
\sigma^2_2 =  ( \sigma^2_{12} + \sigma^2_{23} - \sigma^2_{13} ) / 2 \,, \\[2ex]
\sigma^2_3 =  ( \sigma^2_{13} + \sigma^2_{23} - \sigma^2_{12} ) / 2 \,,
\end{array}
\label{eq:TCH_solution}
\end{equation}
where the $\sigma^2_i$ are the unknown variances of the series and the $\sigma^2_{ij}$ are the observed variances of the paired differences between series.

For an arbitrary number of series $N$, $\sigma_i$ can be found from the following solution \citep{Riley2008}:
\begin{equation}
\begin{array}{l}
\sigma^2_i = \frac{1}{N-2} \left( \sum\limits_{j=1}^N \sigma^2_{ij} - B \right) \,, \\[3ex]
B = \frac{1}{2(N-1)}  \sum\limits_{k=1}^N \sum\limits_{j=1}^N \sigma^2_{kj} \,, \\[3ex]
\sigma_{ii}=0, \ \sigma_{ij}=\sigma_{ji} \,.
\end{array}
\label{eq:Nch_solution}
\end{equation}

This extension of the TCH method is called the N-cornered-hat (NCH) method.

Unfortunately, the method often fails because it may produce negative variances if the data under investigation are correlated.
With correlations, the system to be solved consists of the equations
\begin{equation}
\begin{array}{l}
\sigma^2_{ij} = \sigma^2_i + \sigma^2_j -2 \rho_{ij} \sigma_i \sigma_j \,, \\
\end{array}
\label{eq:nch_corr_equation}
\end{equation}

Accordingly, the key point of the TCH (NCH) method is to find reliable estimates of the correlation coefficients $\rho_{ij}$.


\section{Application of the NCH method to RSPCs}
\label{sect:application}

We used nine recently published RSPCs computed in various VLBI analysis centres.
The general information about these catalogues is given in Table~\ref{tab:catalogs}.
They have 703 sources in common.
All subsequent computations were made for these sources.

\begin{table}
\begin{center}
\caption{Input catalogues (ME is the median position error for the sources in common).}
\label{tab:catalogs}
\begin{tabular}{cclccc}
\hline\hline
Catalogue& Code & Software   &     ME [$\mu$as]    & Note \\
         &      &            & $\alpha$ / $\delta$ &  \\
\hline
aus2012b & AUS  & Occam      & ~76 / ~86 & (1)  \\
bkg2012a & BKG  & Calc/Solve & ~28 / ~40 & (1)  \\
cgs2012a & CGS  & Calc/Solve & ~26 / ~38 & (1)  \\
gsf2012a & GSF  & Calc/Solve & ~24 / ~36 & (2)  \\
igg2012b & IGG  & VieVS      & ~49 / ~62 & (3)  \\
opa2013a & OPA  & Calc/Solve & ~27 / ~37 & (1)  \\
rfc2013a & RFC  & Calc/Solve & 105 / 110 & (4)  \\
sha2012b & SHA  & Calc/Solve & ~27 / ~38 & (3)  \\
usn2012a & USN  & Calc/Solve & ~29 / ~41 & (3)  \\
\hline
\end{tabular}\\
\end{center}
{\bf References.} \\
(1) http://ivscc.gsfc.nasa.gov/products-data/products.html; \\
(2) http://gemini.gsfc.nasa.gov/solutions/astro/; \\
(3) private communication; (4) http://astrogeo.org/rfk/.
\end{table}

First, the paired differences between the catalogues were computed using weights reciprocal to the squares of the source uncertainties.
The differences in right ascension were multiplied by $\cos\delta$.
Then the variances of the paired differences were computed.

The systematic differences between catalogues may have a substantial impact on the determination of the catalogues' stochastic errors.
\citet{Ma2009p} investigated and removed the systematic differences between catalogues prior to evaluating their stochastic errors by means
of the TCH method.
They used a simple model consisting of the rotation and the first harmonic terms, which may be too rough an approximation of
the actual systematics, which is typically much more complicated \citep{Sokolova2007}.
Moreover, the impact of systematic differences on determination of the catalogues stochastic errors was not investigated.

We used a non-parametric representation of the systematic differences between RSPCs by means of exponential smoothing.
We considered this representation preferable for our purpose because a non-parametric model allows one to represent the systematics sufficiently
accurately and without computationally complicated expansion of the catalogue differences into a series of orthogonal functions, which is necessary
to compute a combined catalogue \citep{Sokolova2007}.

The exponential smoothing on the sphere proceeds as follows.
Let a function $y_i$ with the associated uncertainties $s_i$ be given on the celestial sphere for arguments $x_i$ with coordinates
$\alpha_i$ and $\delta_i$, $i=1,\dots,n$.
The smoothed value $y^{*}$ for argument $x^{*}$ with coordinates $\alpha^{*}$ and $\delta^{*}$ is computed by
\begin{equation}
y^{*} = \frac{\sum\limits_{i} {p_i q_i y_i}}{\sum\limits_{i}{p_i q_i}}, \quad q_i = \exp(-d_i^2/2a^2) \,,
\end{equation}
where $p_i$ are weights reciprocal to $s_i^2$, $a$ is the smoothing parameter,
and $d_i$ is the distance between $x_i$ and $x^{*}$ given by
\begin{equation}
\cos d_i = \sin\delta_i \sin\delta^{*} + \cos\delta_i \cos\delta^{*}\cos(\alpha_i - \alpha^{*}) \,.
\end{equation}
The larger $a$, the stronger is smoothing.
Only points with $d/a \le 10$ are included in the sum, which substantially reduces the computation time.
Note that $x^{*}$ must not necessarily coincide with one of the $x_i$.
Hence the method can be used for computing smoothed value for an arbitrary argument, for instance for simultaneous smoothing and interpolation.

\begin{figure*}
\centering
\resizebox{\hsize}{!}{\includegraphics[clip]{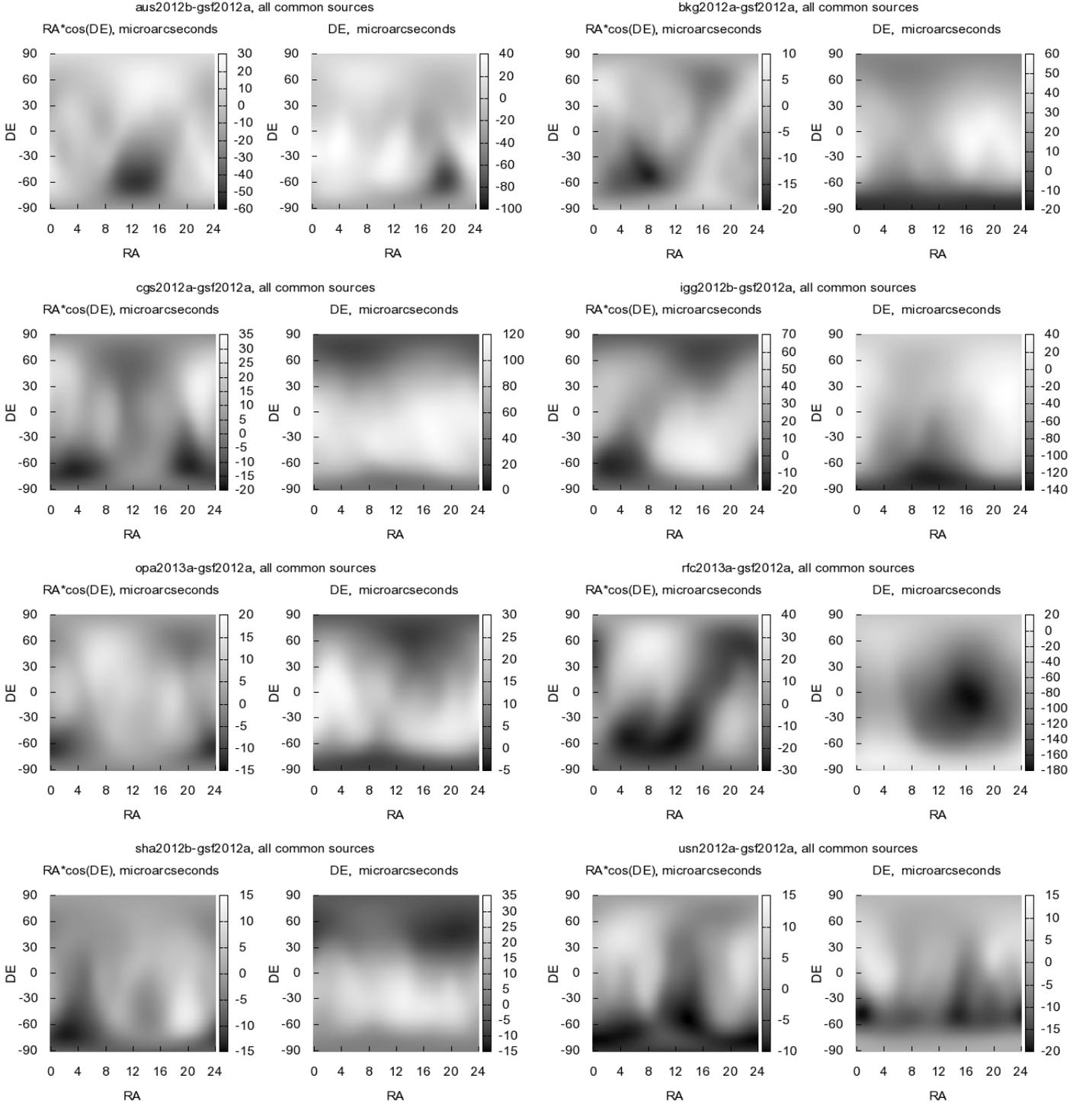}}
\caption{Systematic differences between catalogues.}
\label{fig:systematics}
\end{figure*}

The systematic differences between eight input catalogues and GSF are depicted in Fig~\ref{fig:systematics}.
The systematics in other catalogues' differences can be imagined from these plots.

Although a detailed discussion on the systematic differences between catalogues is beyond the scope of this paper, we can note some remarkable features.
The largest differences can be seen for the RFC catalogue.
One of the main reasons for this may be that the orientation of the RFC catalogue is constrained by 212 ICRF defining sources \citep{Ma1998},
whereas the other catalogues were tied to ICRF2 \citep{Ma2009p}.
ICRF2 was also tied to ICRF, but using the 138 most stable sources common to both ICRF2 and ICRF.
Therefore, a significant part of the systematic errors of ICRF, which reach about 200--250 $\mu$as \citep{Sokolova2007}, can propagate to the RFC catalogue.
However, some other catalogues, computed not with Calc/Solve in the first place, also show large systematic differences both with the GFC catalogues
and among themselves.
This proves that linking individual RSPCs to ICRF2 using the no-net-rotation constraint does not eliminate a significant portion of the catalogue systematic
errors.
Another important conclusion is that the differences in the declination are much greater than in the right ascension, and for most of the differences
a $\Delta\delta_{\delta}$ pattern is clearly visible.
This suggests that different handling of the troposphere (parameterisation, mapping function, gradient modelling, etc.) can be a reason of the
systematics.

To investigate the impact of the systematic differences between catalogues on the result of computing their stochastic errors,
the variances of the paired differences were computed both for the original differences and the differences corrected for the systematics.
These variances are presented in Table~\ref{tab:corr}.

Here, we present a new method for computing the correlations between RSPCs for an arbitrary number of catalogues greater than three.
The proposed computational procedure is as follows:
Let us have $N$ catalogues.
First we select sources in common in all the catalogues, which are used for the subsequent analysis.
In our case we used 703 sources in common for the nine input catalogues.

Now we consider the $i$-th and $j$-th catalogues.
At the first step we computed the differences between these catalogues with all $k$-th catalogues, $k=1,...,N, k \neq i, k \neq j$.
After that, we computed the correlation $\rho^k_{ij}$ between catalogue differences $\Delta_{ik} = Cat_i - Cat_k$
and $\Delta_{jk} = Cat_j - Cat_k$ for each $k$, where $Cat_i$, $Cat_j$, and $Cat_k$ are vectors of the source positions in common.
Computations were made separately for right ascension (RA) and declination (DE).
RA differences were multiplied by $\cos (DE)$.
The average value of $\rho^k_{ij}$ over all $k$ was considered an approximation to the correlation $\rho_{ij}$ between $i$-th and $k$-th catalogues.

To compute the correlation coefficient between two data sets we used both the standard procedure and its weighted modification, defined as follows:
For two series of measurements $x_i$ and $y_i$, $i=1,\ldots,n$, the standard correlation coefficient is computed by
\begin{equation}
\rho_{xy} = \fracd{\sum\limits_{i} (x_i-\bar{x})(y_i-\bar{y})}{\sqrt{\strut\sum\limits_{i} (x_i-\bar{x})^2\sum\limits_{i} (y_i-\bar{y})^2}} \,,
\label{eq:corr}
\end{equation}
where $\bar{x}$ and $\bar{y}$ are the mean values of $x_i$ and $y_i$.

For unevenly weighted series $x_i$ and $y_i$ with associated standard errors $s_{x,i}$ and $s_{y,i}$, we introduced the weighted correlation coefficient as
\begin{equation}
\rho^{w}_{xy} = \fracd{\sum\limits_{i} \sqrt{p_{x,i}p_{y,i}}(x_i-\bar{x})(y_i-\bar{y})}{\sqrt{\sum\limits_{i} p_{x,i}(x_i-\bar{x})^2\sum\limits_{i} p_{y,i}(y_i-\bar{y})^2}} \,,
\label{eq:wcorr}
\end{equation}
where $p_{x,i} = 1/s_{x,i}^2 , p_{y,i} = 1/s_{y,i}^2$, and $\bar{x}$ and $\bar{y}$ are weighted mean of $x_i$ and $y_i$.
Clearly, for evenly weighted data (\ref{eq:wcorr}) is equal to (\ref{eq:corr}).

\begin{figure}
\centering
\resizebox{\hsize}{!}{\includegraphics[clip]{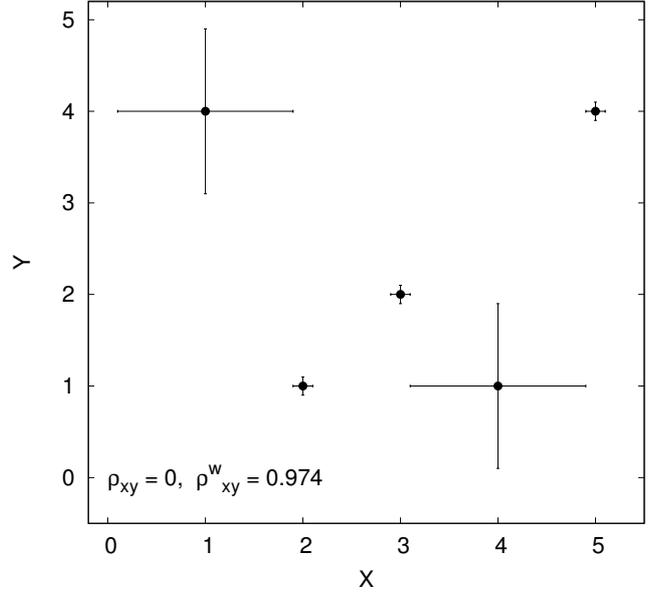}}
\caption{Example of computation of the standard ($\rho$) and weighted ($\rho^{w}$) correlation coefficient for a data with two outliers.}
\label{fig:corr}
\end{figure}

Figure \ref{fig:corr} shows an example of computation of the standard ($\rho_{xy}$) and weighted ($\rho^{w}_{xy}$) correlation coefficient for data
with outliers.
One can see that outliers can lead to a completely wrong correlation estimate.

The correlations between input catalogues and the variances of their paired differences are presented in Table~\ref{tab:corr} for two main variants:
before and after correcting for the systematics.
One can see that the correlations in RA and DE are very similar, and there is no clear dependence on the software.
Evidently, both the variances of paired differences $D$ and the correlation coefficients between catalogues are affected by
the systematic differences between them.
The effect is especially strong for the pairs of catalogues with large systematic differences (see Fig.\ref{fig:systematics}).

\begin{table*}
\centering
\caption{Variances of paired differences ($D$) and correlation coefficients ($\rho$~-- standard, $\rho^w$~-- weighted) between catalogues.}
\label{tab:corr}
\begin{tabular}{ccccccc}
\hline\hline
Catalogues & \multicolumn{3}{c}{Original differences} & \multicolumn{3}{c}{Differences corrected for systematics} \\
           & D [$\mu$as] & $\rho$ & $\rho^w$ & D & $\rho$ & $\rho^w$ \\
           & $\alpha$ / $\delta$ &  $\alpha$ / $\delta$ & $\alpha$ / $\delta$ &  $\alpha$ / $\delta$  & $\alpha$ / $\delta$ &  $\alpha$ / $\delta$ \\
\hline
AUS -- BKG & ~53 / ~67 & $ +0.3981 \ / +0.3108 $ & $ +0.2662 \ / +0.2530 $ & ~50 / ~58 & $ +0.3955 \ / +0.3074 $ & $ +0.2569 \ / +0.2429 $ \\
AUS -- CGS & ~56 / ~73 & $ +0.3249 \ / +0.3266 $ & $ +0.1210 \ / +0.1356 $ & ~50 / ~57 & $ +0.3264 \ / +0.3259 $ & $ +0.1278 \ / +0.1675 $ \\
AUS -- GSF & ~49 / ~58 & $ +0.1858 \ / +0.2027 $ & $ +0.1248 \ / +0.1117 $ & ~46 / ~52 & $ +0.1861 \ / +0.2032 $ & $ +0.1125 \ / +0.1129 $ \\
AUS -- IGG & ~62 / ~76 & $ +0.0541 \ / +0.0518 $ & $ -0.0025 \ / +0.0073 $ & ~50 / ~57 & $ +0.0543 \ / +0.0524 $ & $ +0.0115 \ / +0.0243 $ \\
AUS -- OPA & ~49 / ~59 & $ -0.0741 \ / -0.1048 $ & $ -0.1092 \ / -0.1272 $ & ~45 / ~54 & $ -0.0740 \ / -0.1043 $ & $ -0.0900 \ / -0.0860 $ \\
AUS -- RFC & 106 / 123 & $ -0.3499 \ / -0.1865 $ & $ -0.1697 \ / -0.2005 $ & 102 / 113 & $ -0.3501 \ / -0.1860 $ & $ -0.1536 \ / -0.1529 $ \\
AUS -- SHA & ~48 / ~59 & $ -0.3588 \ / -0.4246 $ & $ -0.2623 \ / -0.3244 $ & ~45 / ~53 & $ -0.3584 \ / -0.4236 $ & $ -0.2207 \ / -0.2397 $ \\
AUS -- USN & ~48 / ~58 & $ -0.6251 \ / -0.7221 $ & $ -0.2661 \ / -0.3505 $ & ~45 / ~52 & $ -0.6264 \ / -0.7263 $ & $ -0.2298 \ / -0.2666 $ \\
BKG -- CGS & ~30 / ~41 & $ +0.6437 \ / +0.6897 $ & $ +0.5553 \ / +0.6449 $ & ~28 / ~33 & $ +0.6428 \ / +0.6907 $ & $ +0.5476 \ / +0.5743 $ \\
BKG -- GSF & ~23 / ~32 & $ +0.5089 \ / +0.6076 $ & $ +0.4824 \ / +0.4897 $ & ~22 / ~28 & $ +0.5082 \ / +0.6095 $ & $ +0.4794 \ / +0.5038 $ \\
BKG -- IGG & ~50 / ~58 & $ +0.0477 \ / +0.1379 $ & $ +0.1332 \ / +0.1968 $ & ~46 / ~50 & $ +0.0481 \ / +0.1374 $ & $ +0.1155 \ / +0.1508 $ \\
BKG -- OPA & ~25 / ~35 & $ +0.1340 \ / +0.1889 $ & $ +0.0712 \ / +0.1080 $ & ~24 / ~32 & $ +0.1335 \ / +0.1892 $ & $ +0.0721 \ / +0.0991 $ \\
BKG -- RFC & ~60 / ~94 & $ -0.2287 \ / -0.0164 $ & $ -0.0464 \ / -0.0009 $ & ~57 / ~71 & $ -0.2286 \ / -0.0165 $ & $ -0.0347 \ / +0.0010 $ \\
BKG -- SHA & ~24 / ~35 & $ -0.2010 \ / -0.1507 $ & $ -0.2268 \ / -0.2196 $ & ~23 / ~30 & $ -0.2013 \ / -0.1515 $ & $ -0.2149 \ / -0.1887 $ \\
BKG -- USN & ~26 / ~38 & $ -0.3170 \ / -0.2214 $ & $ -0.2883 \ / -0.2470 $ & ~25 / ~33 & $ -0.3148 \ / -0.2187 $ & $ -0.2615 \ / -0.1907 $ \\
CGS -- GSF & ~27 / ~46 & $ +0.7685 \ / +0.7713 $ & $ +0.6229 \ / +0.4869 $ & ~23 / ~29 & $ +0.7711 \ / +0.7746 $ & $ +0.6395 \ / +0.6348 $ \\
CGS -- IGG & ~51 / ~70 & $ +0.0246 \ / +0.1577 $ & $ +0.2281 \ / +0.1835 $ & ~46 / ~50 & $ +0.0255 \ / +0.1624 $ & $ +0.2071 \ / +0.2194 $ \\
CGS -- OPA & ~31 / ~45 & $ +0.3127 \ / +0.3160 $ & $ +0.1873 \ / +0.2008 $ & ~27 / ~34 & $ +0.3138 \ / +0.3178 $ & $ +0.2033 \ / +0.2191 $ \\
CGS -- RFC & ~61 / 102 & $ -0.2195 \ / -0.0050 $ & $ +0.0604 \ / +0.0595 $ & ~54 / ~68 & $ -0.2201 \ / -0.0060 $ & $ +0.0780 \ / +0.1045 $ \\
CGS -- SHA & ~27 / ~42 & $ -0.0327 \ / -0.0359 $ & $ -0.0792 \ / -0.0966 $ & ~23 / ~30 & $ -0.0313 \ / -0.0337 $ & $ -0.0624 \ / -0.0499 $ \\
CGS -- USN & ~28 / ~50 & $ -0.1248 \ / -0.1520 $ & $ -0.1423 \ / -0.0627 $ & ~26 / ~33 & $ -0.1229 \ / -0.1531 $ & $ -0.1042 \ / -0.0800 $ \\
GSF -- IGG & ~46 / ~56 & $ -0.0164 \ / +0.2368 $ & $ +0.3538 \ / +0.4202 $ & ~40 / ~44 & $ -0.0193 \ / +0.2334 $ & $ +0.3732 \ / +0.3693 $ \\
GSF -- OPA & ~19 / ~23 & $ +0.5221 \ / +0.4857 $ & $ +0.4751 \ / +0.5206 $ & ~18 / ~21 & $ +0.5210 \ / +0.4823 $ & $ +0.4711 \ / +0.4931 $ \\
GSF -- RFC & ~50 / ~74 & $ -0.2478 \ / -0.0260 $ & $ +0.2164 \ / +0.2849 $ & ~46 / ~55 & $ -0.2497 \ / -0.0311 $ & $ +0.2184 \ / +0.2582 $ \\
GSF -- SHA & ~18 / ~25 & $ +0.1525 \ / +0.1280 $ & $ +0.1186 \ / +0.1195 $ & ~17 / ~20 & $ +0.1528 \ / +0.1270 $ & $ +0.1218 \ / +0.1195 $ \\
GSF -- USN & ~21 / ~24 & $ -0.0271 \ / -0.0832 $ & $ -0.0031 \ / +0.0026 $ & ~20 / ~23 & $ -0.0268 \ / -0.0832 $ & $ +0.0099 \ / +0.0346 $ \\
IGG -- OPA & ~44 / ~57 & $ +0.3165 \ / +0.3643 $ & $ +0.4199 \ / +0.4093 $ & ~39 / ~43 & $ +0.3152 \ / +0.3628 $ & $ +0.4110 \ / +0.4214 $ \\
IGG -- RFC & ~85 / 108 & $ +0.4091 \ / +0.0953 $ & $ +0.2229 \ / +0.2314 $ & ~80 / ~88 & $ +0.4086 \ / +0.0928 $ & $ +0.2482 \ / +0.2508 $ \\
IGG -- SHA & ~47 / ~63 & $ +0.0418 \ / +0.0392 $ & $ +0.1798 \ / +0.0836 $ & ~42 / ~46 & $ +0.0419 \ / +0.0406 $ & $ +0.1907 \ / +0.1805 $ \\
IGG -- USN & ~46 / ~52 & $ -0.0203 \ / +0.0187 $ & $ +0.1318 \ / +0.1188 $ & ~40 / ~42 & $ -0.0211 \ / +0.0177 $ & $ +0.1391 \ / +0.1612 $ \\
OPA -- RFC & ~53 / ~78 & $ +0.0015 \ / +0.2632 $ & $ +0.5582 \ / +0.5127 $ & ~50 / ~59 & $ -0.0003 \ / +0.2607 $ & $ +0.5415 \ / +0.5706 $ \\
OPA -- SHA & ~18 / ~24 & $ +0.6461 \ / +0.6244 $ & $ +0.5367 \ / +0.5367 $ & ~16 / ~22 & $ +0.6479 \ / +0.6239 $ & $ +0.5508 \ / +0.5161 $ \\
OPA -- USN & ~20 / ~26 & $ +0.3423 \ / +0.2585 $ & $ +0.3749 \ / +0.3820 $ & ~18 / ~23 & $ +0.3424 \ / +0.2583 $ & $ +0.3813 \ / +0.4036 $ \\
RFC -- SHA & ~54 / ~82 & $ -0.1097 \ / +0.3237 $ & $ +0.4619 \ / +0.3846 $ & ~50 / ~58 & $ -0.1111 \ / +0.3262 $ & $ +0.5015 \ / +0.5231 $ \\
RFC -- USN & ~63 / 105 & $ +0.1191 \ / +0.0929 $ & $ +0.4034 \ / +0.4213 $ & ~60 / ~94 & $ +0.1194 \ / +0.0907 $ & $ +0.4090 \ / +0.4265 $ \\
SHA -- USN & ~20 / ~29 & $ +0.6927 \ / +0.5710 $ & $ +0.6508 \ / +0.6139 $ & ~19 / ~23 & $ +0.6916 \ / +0.5720 $ & $ +0.6432 \ / +0.6328 $ \\
\hline
\end{tabular}
\end{table*}

We computed the stochastic errors of the nine RFPCs in two ways: with and without correcting for the systematic differences between catalogues.
The results are presented in Table~\ref{tab:errors}.
The weighted correlation coefficients were used in both cases.
A comparison of the two variants shows that the systematic differences significantly affect the determination of their stochastic accuracy.
The numbers in the last column of Table~\ref{tab:errors} are considered as the final result of our work.

\begin{table}
\centering
\caption{Stochastic errors of catalogues found by the NCH method.}
\label{tab:errors}
\begin{tabular}{ccc}
\hline\hline
Catalogue & Original differences & Corrected differences \\
          & $\alpha$ / $\delta$ &  $\alpha$ / $\delta$ \\
          & [$\mu$as] & [$\mu$as] \\
\hline
AUS & 49 / 56 & 46 / 51 \\
BKG & 23 / 27 & 21 / 27 \\
CGS & 27 / 46 & 25 / 27 \\
GSF & 15 / 21 & 14 / 17 \\
IGG & 48 / 59 & 42 / 44 \\
OPA & 15 / 23 & 14 / 18 \\
RFC & 63 / 93 & 60 / 74 \\
SHA & 13 / 17 & 12 / 17 \\
USN & 10 / 12 & 10 / 10 \\
\hline
\end{tabular}\\
\end{table}


\section{Conclusion}
\label{sect_conclusions}

We presented a new approach to assess the external stochastic errors of the RSPCs.
The new features of this method are:

--- simultaneous processing of all catalogues;

--- implementing a new strategy for estimating the correlations between RSPCs;

--- using weighted correlation coefficients;

--- accounting for systematic differences between RSPCs.

Using this approach, we obtained independent estimates of the stochastic errors of the nine recently published RSPCs.
For most of the RSPCs computed in the same manner as their previous versions of 2008, our values generally agree with the estimates obtained
by \citet{Ma2009p}.
For the IGG, RFC, and SHA catalogues, the estimates were computed for the first time.

It is important to note that the external stochastic errors of the RSPCs (Table~\ref{tab:errors}) closely correlate with their formal errors
(Table~\ref{tab:catalogs}).
In other words, we can say that internal and external errors are connected, most probably because of the quality of the software used,
as well as because of analysis strategy details such as modelling and parameterisation.

Indeed, the method developed in this study can be also useful for other catalogues of positions of both celestial and terrestrial objects.

\begin{acknowledgements}
The author is grateful to all the authors of the RSPCs, who made them available to us either via public access (AUS, BKG, CGS, GSF, OPA, RFC)
or via personal contact (IGG, SHA, USN).
The author also thanks the anonymous reviewer for the prompt response and helpful comments.
\end{acknowledgements}

\end{document}